\documentclass
{mn2e}
\usepackage{graphicx}
\usepackage{amsmath, amssymb,bm}

\def \be{\begin{equation}}
\def \ee{\end{equation}}

\def \msun{\rm M_{\odot}}

\def  \le{{L_{\rm Edd}}}

\begin{document}

\title[] { Little Dots: the ULX Analogy}

\author[Andrew King] 
{\parbox{5in}{Andrew King$^{1, 2, 3}$ 
}
\vspace{0.1in} \\ $^1$ School of Physics \& Astronomy, University
of Leicester, Leicester LE1 7RH UK\\ 
$^2$ Astronomical Institute Anton Pannekoek, University of Amsterdam, Science Park 904, NL--1098 XH Amsterdam, The Netherlands \\
$^{3}$ Leiden Observatory, Leiden University, Niels Bohrweg 2, NL--2333 CA Leiden, The Netherlands}
\maketitle
\begin{abstract}
I consider recent observations of the Little Dots (LDs) observed in high--redshift ($z\sim 7$) galaxies. I have suggested that the central black holes in these objects are probably fed mass at very super--Eddington (factors $\sim 50$) rates. In physical terms, this idea makes them
supermassive analogues of the (stellar--mass) ultraluminous X--ray sources (ULXs), whose hard X--ray emission is strongly anisotropic (`beamed'). 

In this paper I argue that the recent discovery of extended rest--frame hard X--ray ionization cones from an LD gives strong quantitative support to this view, as its geometry is what one would expect from a ULX--like object viewed not along one of its X--ray beams, but instead `from the side', very similar to the extreme Galactic system SS433. This in turn agrees in quantitative terms with recent suggestions that LDs resemble supermassive versions of SS433--like systems.

I emphasize that in contrast, the rest--frame {\it soft} X--ray emission from LDs must be roughly {\it isotropic}, and presumably power the torus emission observed from these galaxies.
\end{abstract}



\footnotetext[1]{E-mail: ark@astro.le.ac.uk}
\section{Introduction}

Recent  JWST discoveries of high--redshift 
broad--line (velocity widths a few $10^3\,{\rm km\, s^{-1}}$) AGN showing extreme 
rest--frame X--ray weakness have aroused considerable interest. I shall refer to these objects collectively as 
`Little Dots' or LDs. Some papers distinguish the red systems as LRDs [`Little Red Dots'] from a larger population of similarly compact, broad--line AGN with blue colours [Little Blue Dots (LBDs)]. Estimates in the current literature 
(e.g. Hainline et al. 2025, Taylor et al. 2025, Brazzini et al. 2026, Geris et al. 2026, 
Madau et al. 2026)
find that LRDs constitute approximately 10–30\% of JWST “little dots”, the remaining 70–90\% being bluer. Importantly, both populations appear to share extreme X--ray weakness. This is also consistent with the broader JWST AGN samples, in which X--ray weakness is found well beyond sources classified specifically as LRDs.

In a recent paper (King, 2024; hereafter K24) I suggested that super--Eddington mass 
supply rates\footnote{The commonly--used expression `super--Eddington accretion' is highly ambiguous, as authors frequently employ it to denote either (or both!) of two quite distinct cases: those where the accretor succeeds in swallowing mass at a rate exceeding the Eddington value $\dot M_E$, and others where (as discussed in this paper) mass is supplied at rates $> \dot M_E$, but the accretor rejects a significant fraction of it, and so actually {\it accretes} at rates $\lesssim \dot M_E$.} 
(hereafter SEMS) in high--$z$ quasars (see Fig. 1) and
the resulting high--speed winds carrying away the excess accretion while tightly beaming the emitted rest--frame X--rays, might offer alternative explanations for 
two characteristic features of LD systems. These are (i) the extreme weakness of rest--frame X--ray emission (as we do not in general view them along the X--ray beam), and (ii) the high velocity widths often used instead to argue for large high--$z$ SMBH masses, which are difficult to reconcile with growth by accretion at such early cosmic times.

SEMS and the associated high--speed winds and X--ray beaming are now firmly established (see Lasota \& King, 2023) as the reason why the stellar--mass ultraluminous X--ray sources (ULXs) appear to have hard X--ray luminosities far exceeding any estimate of their relevant Eddington luminosities. Their high--speed ($\sim 0.1c$) winds are observed, and
 ULX systems are now known to be stellar--mass black hole (and sometimes neutron star) binary systems subject to SEMS from a companion star (King \& Lasota, 2023; see King, Lasota \& Middleton, 2023 for a review of ULXs). In (King, 2025) I took this idea further, using results familiar from the study of ULXs to understand observations of high--$z$ supermassive black holes. Since black hole accretion is scale--free, comparing LDs with ULXs is a legitimate procedure. Jun--Rong Liu et al. (2026) take a very similar path, explicitly following the original paper (King et al., 2001) suggesting beaming as the origin of the ULX phenomenon.
 
 The recent paper by Shuying Zhao et al. (2026) follows a similar approach, but 
explicitly models LRDs as supermassive versions not simply of ULXs, but of extreme stellar--mass binary systems like SS433, which is now known (Begelman, King \& Pringle 2006) to be a ULX system viewed `from the side', that is, not down the X--ray beam.
 
 In this paper I will discuss the two characteristic features of SEMS known from ULXs, and apply these results to LDs. These two features are 

(i) a quasispherical gas outflow driven by radiation pressure with speeds $\sim 0.1c$, emitting a roughly isotropic luminosity $\sim \le$ in rest--frame soft X--rays, and

(ii) strongly anisotropic (`beamed') rest--frame hard X--ray emission, which results because angular momentum conservation for the gas expelled in the quasispherical outflow requires it to leave a double radial vacuum channel of small solid angle.

The characteristic $\sim 0.1c$ velocity arises because this is the escape value from the black--hole ISCO. This velocity is directly inferred from observations of the extreme stellar--mass binary SS433, which is so far the only fairly clear case of a ULX seen `from the side' (Begelman, King \& Pringle, 2006) rather than along its X--ray beam. (Begelman et al. 2006 discuss why SS433 is our Galaxy's expected `ration' for super--Eddington accretors, whereas stellar--mass ULXs are found from observations of a far larger sample of distant galaxies.)

The radial channels noted in (ii) produce the characteristic ULX property: scattering within the outflowing wind means that much of the centrally emitted accretion luminosity finds these channels and ultimately escapes along them, making the radiation specific intensity viewed along each channel very high. To a distant observer within a beam but unaware of its strongly anisotropic nature this implies a total luminosity significantly exceeding the Eddington value $L_E$, hence the `ultraluminous' description.

In Section 2 I show that there is evidence favouring the idea that high--$z$ supermassive black holes (particularly in LDs) are subject to SEMS, and  are high--mass analogues of ULX--type 
systems which, like SS433, are oriented with their rest--frame X--ray beams away from the line of sight. 

 In Sections 3 and 4 I discuss the consequences of features (i) and (ii) respectively, particularly in the light of the recent discovery by Zhiyuan Li et al. (2026; hereafter ZL26) of ionization cones in a galaxy hosting an LD. This gives independent quantitative support to the idea that 

I will also show explicitly that the the resulting beaming in these systems is never total, i.e. that the strongly anisotropic rest--frame X--ray component is alway accompanied by a much softer {\it near--isotropic} component which powers the emission from the dusty torus.

\section{Super--Eddington Mass Supply}

If LD black holes are supplied with mass at super--Eddington rates, the resulting high--speed near--isotropic expulsion (cf (i) above) of most of this gas explains why many high--$z$ SMBH show emission lines with velocity widths a few $1000\, {\rm km\,s}^{-1}$. If these widths are
instead interpreted as coming from gas on {\it bound} orbits about the black hole this requires
the existence of very high SMBH masses at high $z$, which are difficult to explain at such early cosmic times.

The suggested presence of these near--isotropic winds makes LDs closely analogous to the ULXs observed in the low--redshift Universe. ULXs are an evolutionary phase of high (stellar) mass X--ray binaries (see King, Lasota \& Middleton 2023 for a review). Here stellar--mass black holes (or sometimes neutron stars) are fed mass at highly super--Eddington rates by their binary companion stars\footnote{This happens because the donor star is more massive than the black--hole or neutron--star accretor. Since mass transfer then puts gas further from the binary centre of mass, the binary separation must shrink in order to conserve angular momentum, reinforcing the tendency of the donor to overfill its Roche lobe and so increasing the mass transfer rate. This is ultimately limited only by the thermal timescale on which the donor star can heat and expand the
lower--entropy gas rising from its core to maintain its surface radius and continue filling its Roche lobe (see King et al.,  2023, Section 3.11).}. 
LD black holes must expel much of this gas at the same characteristic velocity $\sim 0.1c$ seen in ULXs, since black--hole accretion is scale--free. ULX winds are now known to collimate (`beam') a significant fraction ($\sim 50$~\%) of the accretion luminosity (emitted in rest--frame X--rays) into a double radial cone of small solid angle $\sim \pi/10$ around the mean angular momentum vector of the accreting gas. This beaming accounts for the `ULX' property, as an observer within the beam, but unaware of it and instead assuming spherical symmetry, would assign a much larger apparent luminosity to the source. Clearly most ULX systems must actually be unrecognised, as we can generally identify only those where one of the double X--ray beams points at us. 

Significantly,  ZL26  recently found an LD system with an ionization cone on the sky consistent with beamed emission in rest--frame medium--energy X--rays directed across our line of sight. I will show that the opening angle of these cones agrees with the quantitative estimate of the beam opening angle from ULX physics, given the expected Eddington factor of the accreting gas.
I argue that super--Eddington mass supply to 
$\sim 10^8\msun$ SMBH and the consequent beamed emission appears to offer a valid explanation for observations of LD galaxies at $z \sim 7$, making LDs supermassive analogues of ULXs. 
%
%

I first summarise the current understanding of SEMS in the context of ULXs, where it is more complete (see the review by King, Lasota \& Middleton, 2023 for more details).

In ULXs the accretor (a stellar--mass black hole, or neutron star) 
reacts to a strongly super--Eddington mass supply rate 
$\dot M \gg \dot M_E = L_E/\eta c^2$ (where
$L_E = 4\pi GMc/\kappa$ is the Eddington luminosity,
with $\kappa$ the electron--scattering opacity and $\eta \sim 0.1$ the accretion efficiency) from its companion star
by progressively expelling the excess (i.e locally super--Eddington) inspiralling gas from each radius of the accretion disc (Shakura \& Sunyaev, 1973). This produces a roughly spherical high--speed 
 \be
 v \sim \frac{0.1c}{\dot m} 
 \ee
Compton--thick outflow, where $\dot m = \dot M/\dot M_E \gg 1$, with $\dot M$ the mass supply rate. The accretor actually gains mass only at a rate $\sim \dot M_E \ll \dot M$,  as each annulus of the disc adjusts its local accretion rate to the local radiation--pressure limit by expelling the supercritical excess, producing the outflow. This 
 solution is detailed in the original paper by Shakura \& Sunyaev (1973), who showed that the inward accretion rate through the disc varies as $\dot M(R) \sim \dot M_E (R/R_{\rm in})$, where $R_{\rm in}$ is the inner disc radius specified by the black hole ISCO. 
 (For numerical treatments and further discussion see e.g. Ohsuga et al., 2005; King, Lasota \& Middleton 2023; King, 2023.)
 
This process leads to a total luminous output
 \be
 L \simeq \le (1 + \ln\dot m).
 \ee
 The wind carrying off the excess accretion is roughly spherical, but angular momentum conservation requires two narrow vacuum cones within it, along the rotational axis of the central disc gas, as none of the accreting gas has its angular momentum entirely removed. A component 
 $\sim \le$ of the total accretion luminosity (both directly emitted, and scattered by the outflow) finds these cones and escapes to infinity: the narrow solid angles of the cones mean that the radiation components along them have a very high specific intensity.

A similar luminosity $\sim \le$ diffuses outwards through the quasispherical outflow, and is emitted roughly isotropically from its photosphere (where it shocks against the ISM of the host galaxy) as significantly softer radiation, with far lower specific intensity than the beamed emission.

One can deduce (King, 2009) that this picture leads to intense X--ray emission in a double beam of solid angle $4\pi b \ll 1$, with 
\be
b \simeq\frac{73}{\dot m^2},
\label{b}
\ee
where as before $\dot m = \dot M/\dot M_E$. An observer located within either of the beams and assuming spherical emission would deduce an apparent luminosity 
\be
L \simeq \frac{1}{b}\le \gg \le
\label{app}
\ee
in X--rays. This picture is now widely accepted as the likely origin of the ultraluminous X--ray sources. In particular, recent observational evidence (Veledina et al., 2024; see also Lasota \& King, 2023) gives explicit support to it, in that the well--known stellar--mass X--ray binary 
Cyg~X--3 is revealed through X--ray polarization data to have a powerful energy stream oriented orthogonal to the line of sight, and so is a ULX `seen from the side'.

Figure 1 shows a schematic picture of the accretion geometry produced by a super--Eddington mass supply to a black hole. For a stellar--mass system there is no additional structure (apart from the companion star supplying the accreting gas), but if the accretor is a high redshift SMBH, the structure shown would be surrounded by an AGN torus resulting from the interaction with the host galaxy. This is omitted from Fig. 1 for clarity.
\begin{figure}
\hspace*{-1cm} 
     \centering
     \includegraphics[width=0.8\columnwidth, angle=0]
     {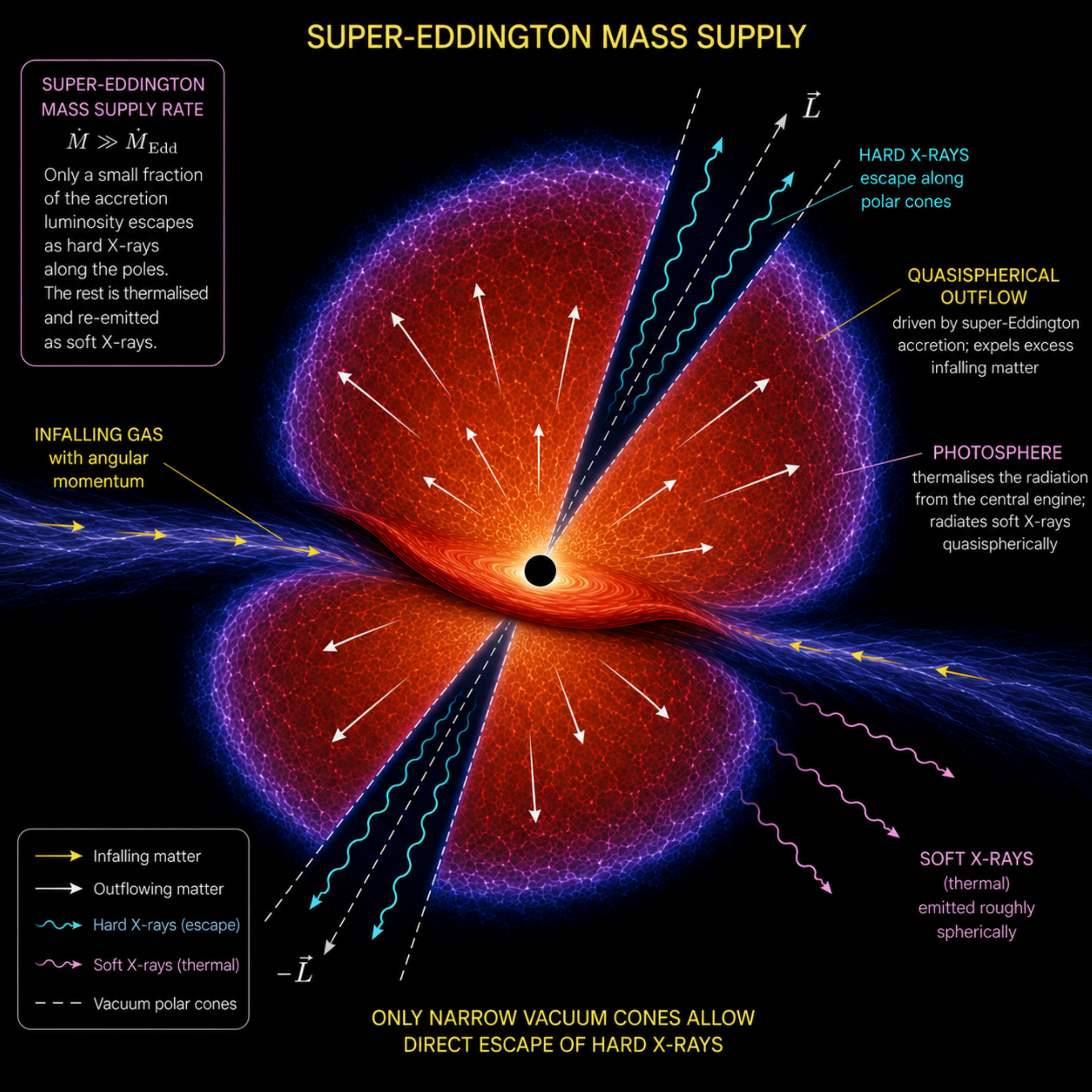}
     \caption{Schematic picture of super--Eddington mass supply to a black hole, and the consequent differing radiation patterns for hard and soft X--ray emission.   
The excess accreting gas is ejected quasispherically by radiation pressure, leaving two narrow vacuum channels along the rotational axis of the accretion flow close to the hole, given by the angular momentum of the innermost disc. Hard--to--medium rest--frame X--ray emission produced near the hole finds these channels after scattering, and escapes to infinity in a narrow double cone: it is only this emission which is beamed. A softer radiation component is emitted quasi--isotropically from the photosphere of the outflow where it shocks against the host ISM.
If the hole is supermassive the structure shown here is surrounded by a dusty torus (not shown, for clarity) whose structure is determined by interaction with the larger--scale host galaxy.  The very narrow solid angles of the (rest--frame) hard X--ray beams imply that in a small sample of 
high--redshift galaxies it is relatively unlikely that their tori would intercept them. Accordingly any torus `sees' only the same roughly isotropic rest--frame soft X--ray radiation pattern as a distant observer.  This picture is generic: in the present paper the hole is assumed to be supermassive, i.e. with $10^5 \lesssim M/\msun \lesssim 10^8$.  
For low--redshift stellar--mass black holes (or neutron stars) with super--Eddington mass supply rates this same picture gives a widely accepted picture of the accreting component of a ULX (King et al., 2023) if the viewer lies in one of the hard X--ray emission cones. From all other viewpoints, the system appears similar to the extreme Galactic stellar--mass binary system SS433.}
\label{fig.1}
\end{figure}

Despite their similar structure there is one important observational difference between the ULX and LD cases. In ULXs the origin of the super--Eddington mass supply is a single lobe--filling star orbiting the black hole in an X--ray binary system. The observer must be at a particular orientation to this orbit  -- in the beam -- in order to recognise the ultraluminous property. 

In contrast, in LDs, SEMS events presumably occur at several separate epochs and at varying orientations, because accretion is driven by the chaotic interaction with neighboring galaxies. This property will be important in understanding the observation of ionization cones from an LD.

\section{Anisotropic Emission of Rest--Frame X--rays in High--Redshift AGN}

A straightforward line of reasoning supports the idea that many 
high--redshift AGN are undergoing SEMS, and so emit anisotropically to some degree. 
High--$z$ SMBH inhabit a densely--packed region of the early Universe. It is natural to expect the gravitational interactions between their hosts and other galaxies to drive gas accretion on to many of them at rates close to the dynamical value 
\be
\dot M_{\rm dyn} \sim \frac{f_g \sigma^3}{G}, 
\label{dyn}
\ee
where $f_g$ is the gas fraction wrt dark matter. The Salpeter time is shorter than the age of the Universe for $z \lesssim 7$, so SMBHs at such redshifts will have reached masses $M$ close to the $M - \sigma$ relation (cf King, 2003; 2005), i.e.
\be
M \simeq M_{\sigma} \simeq \frac{f_g\kappa}{\pi G^2}\sigma^4,
\label{msig}
\ee
where $f_g$ is the gas fraction relative to all matter, and $\kappa \simeq 0.34{\rm cm}^2/{\rm g}^{-1}$ is the electron scattering opacity. This implies an Eddington ratio
\be
\dot m \sim \frac{\dot M_{\rm dyn}}{\dot M_E} \sim \frac{54}{ M_8^{1/4}},
\label{dyn}
\ee
where $M_8 = M/10^8\msun$,
suggesting that many of these systems undergo SEMS. 

From (\ref{b}), equation (\ref{dyn}) implies a typical beaming factor $b \sim M_8^{1/2}/34$ in LDs. With $M_8 \sim 1$ this is in good agreement with the finding (Maiolino et al., 2024) that only 2 LD systems from a sample of 71 are detectable in rest--frame X--rays. 

\section{Ratio of `ULX--like' to `SS433-like'}

The basis of the present paper is the distinction between SMBH which appear like ULXs, and those (the LDs) which appear like supermassive SS433 systems instead. For stellar--mass binary systems the disparity in detections of these prototypes is very clear: there are currently approaching 2000 known ULX systems (see Middleton et al., 2023 and references therein), and still only one clear case of an SS433--like system (SS433 itself), with a handful ($3 - 4$) of possible further candidates identified from jet--inflated nebulae. But as we have seen, the situation with SMBH systems is effectively the complete opposite: all but 2 of the 71 current LD sample have the characteristics of SS433, i.e. weak or absent rest--frame hard X--rays. So there are about 35.5 as many SS433--type systems as ULX--type ones.

The explanation for this wide disparity must come from the typical beaming factor $b$ -- we only detect a fraction $b$ of ULX--like systems of a given luminosity, so the total number of ULX--like systems is greater than the observed number by a factor $1/b$. Conversely, beaming has
little effect on finding SS433--like systems, except for removing a few -- we see a fraction $1-b$ of those luminous enough to see. 
Assuming that the typical Eddington luminosity for the two samples is similar (i.e. there is no systematic difference in SMBH masses) in the two types, the ratio of detectable systems is
\be 
\frac{\rm ULX}{\rm SS433} \simeq \frac{1 - b}{b}  \sim 35.5,
\label{ul/ss}
\ee
which gives $b^{-1} \sim 36.5$. From the standard ULX beaming formula (\ref{b}) the estimate
(\ref{dyn}) successfully predicts this ratio.

\section{Ionization Cones in LDs}


Elementary geometry predicts the aspect ratio (width/length) of a typical observed ionization cone for ULX--like systems as  $r/R \sim 2b^{1/2}\sim 0.33$, up to geometrical projection factors $\sim 1$.



Then the definition of $b$ constrains the square of the relative width--to--length ratio of the cones as 
\be
\frac{r^2}{R^2} = 4b \simeq 0.1 M_8^{1/2},
\ee
and so requires the angular width of the ionization cone on the sky to be 
about 1/3 of its length. 

This is reasonable when compared with Fig.~1 of Zhiyuan Li et al. (2026).

.

\section{Conclusion}

Super--Eddington mass supply on to SMBH of masses $\sim 10^8\msun$ remains 
a valid explanation for observations of active galaxies at $z \sim 7$, in particular the Little Dots.
These appear to be SS433--like systems, where moderate--mass SMBH accrete at the 
super--Eddington mass rates expected from dynamical interaction between their host galaxies, and are oriented with the rest--frame X--ray emission cones away from the line of sight. The observation by Zhiyuan Li et al. (2026) of the geometry of ionization cones in an LD gives an
independent test of this picture.
 
There is a close analogy between these systems and many aspects of stellar--mass ultraluminous X--ray sources (ULXs). It may be possible to test these ideas directly by further observations of the structure found by Zhiyuan Li et al. (2026), and future similar discoveries.

\section*{DATA AVAILABILITY}
No new data were generated or analysed in support of this research.

\section*{ACKNOWLEDGMENTS}
I thank the referee for comments which greatly improved this paper. The schematic picture of super--Eddington mass supply was generated with the assistance of Open AI Chat GPT and checked by the author.

{}

\appendix

\section{The Meaning of Beaming}
A recent paper (Bosman et al., 2025) claimed that the suggested beaming (K24) of a significant component of the accretion luminosity in high--redshift AGN is ruled out.
The stated reason was that beamed emission from the SMBH could have no effect on its immediate surroundings outside the beam direction, and so would be unable to power the emission from the dusty torus. This would contradict the observed
rest-frame optical/IR properties of the four highest-redshift known luminous type~1 quasars at $7.08 < z < 7.64$, which indicate reprocessing of substantial emission from the black hole's vicinity. 

I contend that the claim by Bosman et al. (2025) is not correct, because it implicitly assumes that the beaming suggested in K24 is {\it total}, in the sense that  {\it all} of the accretion luminosity 
$\simeq \le(1 + \ln\dot m)$ 
of the central SMBH region is emitted in a narrow solid angle, and there is no wide--angle emission outside this narrow beam. 
 
But K24 did not propose this, and instead explicitly stated that the beaming predicted by this picture is partial, in the sense that a significant luminosity component must also be emitted 
{\it quasi--isotropically} from the photosphere of the outflowing expelled gas (cf Fig. 1).
The rest--frame {\it soft} X--ray/UV emission from these high--$z$ AGN is roughly isotropic, 
since the quasispherical photosphere of the expelled gas flow (solid angle $\sim 4\pi$) is heated to temperatures $\simeq 10^{4 - 5}$~K. This much softer isotropic emission heats the AGN torus surrounding the nucleus, whose orientation is determined by the large--scale structure of the galaxy. This directly contradicts the argument in Bosman et al. (2025), claiming to rule out significantly anisotropic (`beamed') emission from the vicinity of the central SMBHs in high--redshift quasars. As demonstrated above (see also King \& Muldrew, 2016), this soft component also has total luminosity 
$\sim L_E$. 
 
Equation (19) of King \& Muldrew (2016) shows that the effective temperature of this component is
\be
T_{\rm eff} \sim 5\times 10^4~{\rm K},
\ee
which is of order typical values for an AGN torus.

K24 further noted that this kind of partial beaming was perfectly compatible with the paper by Haiman \& Cen (2002), which had already shown that quasar emission could not be totally beamed towards the observer.

\end{document}